\documentclass[12pt]{article}
\usepackage{rotate}
\usepackage{epsfig}
\usepackage{psfrag}
\usepackage{a4}
\usepackage{color}

\newcommand \be  {\begin{equation}}
\newcommand \bea {\begin{eqnarray} \nonumber }
\newcommand \ee  {\end{equation}}
\newcommand \eea {\end{eqnarray}}

\begin{document}

\title{A proposal for impact-adjusted valuation: Critical leverage and execution risk}

\author{F. Caccioli$^{1}$, J.-P. Bouchaud$^{2}$, J.-D. Farmer$^{1}$
~~~~~\\
{\em 1 - Santa Fe Institute, 1399 Hyde Park road, Santa Fe, NM 87501, USA }\\
{\em 2 - Capital Fund Management, 6-8 Boulevard Haussmann, 75009 Paris, France}\\ 
}

\maketitle

\begin{abstract}
The practice of valuation by marking-to-market with current trading prices is seriously flawed.   Under leverage the problem is particularly dramatic:  due to the concave form of market impact, selling always initially causes the expected leverage to increase.   There is a critical leverage above which it is impossible to exit a portfolio without leverage going to infinity and bankruptcy becoming likely.  Standard risk-management methods give no warning of this problem, which easily occurs for aggressively leveraged positions in illiquid markets.  We propose an alternative accounting procedure based on the estimated market impact of liquidation that removes the illusion of profit. This should curb the leverage cycle and contribute to an enhanced stability of financial markets.

\end{abstract}

\vspace{.2in}
{\it Another issue brought to the fore by the crisis is the need to better understand
the determinants of liquidity in financial markets. The notion that
financial assets can always be sold at prices close to their fundamental
values is built into most economic analysis...}

Chairman Ben Bernanke, {\it Implications of the Financial Crisis for Economics}, Princeton, September 24th, 2010.

\section{Introduction:  The danger of marginal prices}

Mark-to-market or ``fair value'' accounting is standard industry practice. It consists in assigning a value to a position held in a financial instrument based on the current market price for this instrument or similar instruments.   This is justified by the theory of efficient markets, which posits that at any given time market prices faithfully reflect all known information about the value of an asset.  However, mark-to-market prices are only {\it marginal} prices, reflecting the value of selling an infinitesimal number of shares.  

Practitioners are typically concerned with selling more than an infinitesimal number of shares, and are intuitively aware that this practice is flawed.  Selling has market impact, which depresses the price by an amount that increases with the quantity sold.  The first piece will be sold near the current price, but as more is liquidated prices may drop substantially.  This somewhat paradoxically implies the value of 10 \% of a company is less than 10 times the value of 1 \% of that company.  We take advantage of what has been learned recently about market impact to propose a method for {\it impact-adjusted valuation} that results in better risk control than mark-to-market valuation.  This is in line with other recent proposals that valuation should be based on liquidation prices \cite{Acerbi08,Caccioli11}.

Estimating liquidation prices requires a good understanding of market impact.  In recent years there is been considerable progress in both theory and practice.   For large trades there is growing evidence that market impact follows a universal functional form, see e.g. \cite{Torre97,Almgren05,Toth11b}.  By ``large" we mean trades that exceed the liquidity currently available in the order book; such trades need to be either broken up into pieces and executed incrementally or executed in a block market.  Market impact is a concave function whose slope is infinite at the origin, which means that small trades have a disproportionally large impact. 

The need for a better alternative to marking to market is most evident with leverage, i.e. when assets are purchased with borrowed money.  Leverage amplifies market impact.  As a leveraged position is sold, a process we refer to here as {\it deleveraging}, the price tends to drop due to market impact.   Counter-intuitively, due to the concave form of market impact, when a leveraged position is gradually unwound the depression in prices due to impact overwhelms the decrease in position size, and leverage initially rises rather than falls.  When impact is concave, this is not unusual -- the expected leverage as a sale begins always goes up, regardless of initial leverage, liquidity or position size.

As more of the position is sold, provided the initial leverage and initial position are not too large, leverage eventually comes back down and the position retains some of its value upon liquidation.  However, as we show here, if the initial leverage and initial position are too large, as the position is sold the leverage diverges, and the resulting liquidation value is less than zero, i.e.  the debt to the creditors exceeds the resale value of the asset.  The upshot is that under mark-to-market accounting a leveraged position that appears to be worth billions of dollars may predictably be worth less than nothing by the time it is liquidated.   The above scenario assumes that positions are exited in an orderly fashion; under fire sale conditions or in very illiquid markets things are even worse.  

From the point of view of a regulator or a risk manager this makes it clear that an alternative to mark-to-market accounting is badly needed.  Neglecting impact allows huge positions on illiquid instruments to appear profitable when it is actually not the case. We propose such an alternative based on the known functional form of market impact, and propose that valuations should be based on the expected liquidation value of assets.  Under leverage this avoids the problems outlined above.  Whereas mark-to-market valuation only indicates problems with excessive leverage after they have occurred, our method makes them clear before positions are entered.  Thus our method gives clear indications about potential problems as they are developing, and makes such situations easier to avoid.  This could be extremely useful for damping the leverage cycle and coping with pro-cyclical behaviors \cite{Geanakoplos03,Adrian09,Brunnermeier09,Thurner09,Geanakoplos10}.

In Section 2 we review the literature on market impact and present our proposal for impacted-adjusted valuation.  In Section 3 we apply this to leverage and demonstrate that over-leveraging is a critical phenomenon, with a sharp transition where the problem of liquidating the position without bankruptcy becomes serious.  In Section 4 we present some alternative formulas for estimating impact, run some numbers for typical assets, and show that this is not a serious problem for really liquid assets such as stocks, but it can occur at surprisingly low leverages for assets such as credit default swaps.  Section 5 concludes, discussing the broader implications for the theory of market efficiency.

\section{Market impact and liquidation accounting}

Accounting based on liquidation prices requires a quantitative model of market impact.  Because market impact is very noisy, and because it usually requires proprietary data to study empirically, a good picture of market impact has emerged only gradually in the literature.  In this section we review what is known about market impact and present our proposal for impact-adjusted valuation.

\subsection{The emerging quantitative model of market impact}

Understanding the nature of market impact has now been the focus of a large number of empirical studies, both from academics and practitioners (for recent reviews, see \cite{Bouchaud08b,Bouchaud10,Torre97, Almgren05, Engle08,Moro09,Toth11b}), and a consensus is beginning to emerge. Here we are particularly concerned with the liquidation of large positions, which must either be sold in a block market or broken into pieces and executed incrementally\footnote{ Our interest in the impact of a single large trade that must be executed in pieces is in contrast to the impact of a single small trade in the orderbook, or the impact of the average order flow, both of which have different functional forms, see \cite{Bouchaud10,Toth11b}.}.
These empirical studies now make it clear that the market impact $I = \langle \varepsilon \cdot (p_{f} - p_{0})/p_0 \rangle$, 
defined as the expected shift in price from the price $p_0$ observed before a buy trade 
($\varepsilon=+1$) or a sell trade ($\varepsilon=-1$) to the price $p_f$ at which the last share is executed, is a concave function of position size $Q$ normalized by the trading volume $V$.   When liquidation occurs in normal conditions, i.e. at a reasonable pace that does not attempt to remove liquidity too quickly from the order book, the expected impact $I$ due to liquidating $Q$ shares is
\be\label{impactQ}
{I}(Q) = Y \sigma \, \sqrt{\frac{Q}{V}},
\ee
where $\sigma$ is the daily volatility, $V$ is daily share transaction volume, and $Y$ a numerical constant of order unity \cite{Toth11b}.   We say more about how these parameters should be estimated when this formula is used for regulatory purposes in the next section.

Note that we are defining the expected impact in terms of prices rather than log-prices.  This is possible because for cases of interest the liquidation time $T$ is short enough for prices not to move significantly away from the initial price, and the impact itself is significantly less than the price itself, so that the difference between $p_f/p_0 - 1$ and $\ln(p_f/p_0)$ is small and only has a minor effect on our conclusions.  This means that the domain of validity for the formula requires that the impact $I(Q)$ not be too large, roughly less than $20\%$.

The quantity above is the expected impact, in the sense that it is the average outcome of liquidating $Q$ shares.  This is superimposed on the background price fluctuations due to the rest of the market.  For typical small values of $Q/V$ allowing orderly execution, the realized market impact is very noisy, almost invisible to the naked eye.  It is not uncommon that the realized impact is in the opposite direction of the average impact.  The expected impact can be regarded either as the average impact or as a median price -- $50\%$ of prices will be above it, and $50\%$ below it.

We want to emphasize that here we have defined impact as the shift in prices caused by the execution of given order.  Whether the long-term impact has a permanent component that remains embedded in prices long after the trade occurs, and how large such a component might be, remain controversial.  Fortunately these questions do not need to be addressed for our purposes here, although they are highly relevant to understand
how market prices move and how potentially destabilising feedback loops can occur (see e.g \cite{JP_Risk}).

The earliest theory of market impact due to Kyle \cite{Kyle85} predicted that expected impact should be linear.  This was further supported by the work of Huberman and Stanzl \cite{Huberman04b}, who argued that providing certain assumptions are met, such as lack of correlation in order flow, impact has to be linear in order to avoid arbitrage.  However, more recent empirical studies have made it clear that these assumptions are not met \cite{Bouchaud04,Lillo03c,Toth11b}, and the overwhelming empirical evidence that impact is concave has driven the development of alternative theories \cite{Grinold95,Gabaix06,Gatheral08}.  For example, Farmer et al. \cite{Farmer11} have proposed a theory based on a strategic equilibrium between liquidity demanders and liquidity providers, in which uncertainty about $Q$ on the part of liquidity providers dictates the functional form of the impact.  Toth et al. \cite{Toth11b}, in contrast, derive a square root impact function within a stochastic order flow model.  They impose that prices are diffusive, show that this implies a locally linear ``latent order book'', and provide a proof-of-principle using a simple agent-based model.  Both of these theories predict roughly square root impact, though with some differences.

Both empirical studies and theory make it clear that the square-root law for expected impact under orderly execution also holds at intermediate points.   That is, after a quantity $q \leq Q$ is executed, the average adverse price move is given by \cite{Moro09,Farmer11,Toth11b}:
\be
{I}(q) = Y \sigma \, \sqrt{\frac{q}{V}}, 
\ee

We should stress that the formulae above for market impact hold only under normal conditions, when execution is slow enough for the order book to replenish between successive trades (on this point, see e.g. \cite{Weber05,Bouchaud04b,Bouchaud08b}). If the execution schedule is so aggressive that $Q$ becomes comparable to $V$, liquidity may dry up, in which case the parameters $\sigma$ and $V$ can no longer be considered to be fixed, but themselves react to the trade, with an expected increase of the volatility and a decrease of the liquidity. Impact in such extreme conditions is expected to be much larger than the square-root formula above.  The flash crash is a good example.  In these cases the expected impact becomes less concave and it can become linear or even super-linear \cite{Gatheral08}.
For the above impact formula to be valid, the execution time $T$ needs to be large enough that $Q$ remains much smaller than $V$ ($20\%$ is a typical upper limit). The execution time should not be too long either, otherwise impact 
is necessarily linear in $Q$: beyond some ``memory time'' of the market, trades must necessarily become independent and impact must be additive, see \cite{Toth11b}. 

 \subsection{How should the impact parameters be estimated?}
 
When impact is estimated for regulatory purposes, for stability reasons it is important that the parameters should be computed over a long time horizon.  For example one can take an exponential moving average of $\sigma$ and $V$ over past values. If $\sigma$ and $V$ are not measured over relatively long time scales impact-adjusted valuation could lead to an unstable feedback loop.  Imagine, for example, an exogeneous shock (like the Japanese tsunami in March 2011) that leads to a sudden increase of volatility.  If $\sigma$ is measured over short-time scales, the
expected impact $I(Q)$ also increases.  This would cause a larger discount on the asset valuation, which could cause a systemic effect in which risk managers unload the asset, leading to plummeting market prices and further panic. Similarly in a temporary liquidity crisis a sudden drop of $V$ could lead to a mechanical reduction in asset values. In order to avoid these destabilising effects, 
the window over which $\sigma$ and $V$ are computed should be chosen to be long, perhaps 6 months, and exclude the very recent past -- e.g. the last week of trading.

\subsection{Impact-adjusted accounting}

The establishment of a quantitative theory for expected impact makes it possible to do impact-adjusted accounting.  Rather than using the mark-to-market price, which is the marginal price of an infinitesimal liquidation, we propose using the expected price under complete liquidation.  For convenience we assume liquidation in $N$ equal sized increments of $v$ shares each, where $v$ is arbitrary but small\footnote{
In general it is possible that optimal liquidation might follow a different liquidation schedule.  However, our feeling is that any gains from such a schedule are likely to be small, and in any case, empirical studies show that a uniform liquidation rate is a good approximation for the average investor \cite{Moro09}.}.
The expected value $\mathcal{V}$ of a position of $Q$ shares in a given asset with mark-to-market price $p_0$ that is liquidated in $N$ pieces of $v$ shares each is 
\be
\mathcal{V} (Q) = \sum_{t=1}^N v p_0 (1 - I(vt))
\ee
Providing $Q$ is large, it is a good approximation to use the continuous limit where $dq = v$ is infinitesimal, in which case this can be written
\begin{eqnarray}
\mathcal{V} (Q) & = & \int_0^{Q}  p_0 (1 - I(q)) dq\\
\nonumber
& = & p_0 Q(1 - \frac{2}{3} Y \sigma \sqrt{Q/ V})\\
\nonumber
& = & p_0 Q(1 - \frac{2}{3} I(Q))
\end{eqnarray}
 It is sometimes also useful to use the {\it average valuation price} $\tilde{p} = \mathcal{V}/Q = p_0 (1 - \frac{2}{3} I(Q))$.

\section{The critical nature of leverage}

When leverage is used it becomes particularly important to take impact into account and value assets based on their expected liquidation prices.  Consider an asset manager taking on liabilities $L$ to hold $Q$ shares of an asset with price $p$.  For simplicity we consider the case of a single asset.  The leverage $\lambda$ is given by the ratio of the value of the asset to the total equity,
\be\label{before}
\lambda=\frac{Qp}{Qp-L}.
\ee 
Holding $Q$ and $L$ constant, the leverage decreases when the price of the asset increases and vice versa when it decreases.  Similarly, holding $p$ and $L$ constant, selling $q$ shares reduces leverage,
\be\label{before2}
\lambda=\frac{Qp}{Qp-L} \to \lambda' = \frac{(Q-q)p}{(Q - q)p-L} < \lambda \quad {\mbox {if}}\quad  q > 0,
\ee
and vice versa for buying.

\subsection{Deleveraging}

Now we take into account market impact and consider the case of deleveraging, i.e. exiting a leveraged position.  Selling pushes current trading prices down, which under mark-to-market accounting decreases the value of the remaining unsold shares. As we show, this generally overwhelms the effect of selling the shares, increasing the leverage even as the overall position is reduced.   
After $q$ shares have been sold the amount of cash raised to offset the liabilites is $C(q)$.  Using the continuous approximation
\be
C(q) \approx  \int_0^q \,{\rm{d}}q' \, p_0 \left(1 - I(q')\right) = p_0 q \left(1 - \frac23 {\cal I}  \sqrt{\frac{q}{Q}}\right),
\ee
where ${\cal I} \equiv I(Q)= Y \sigma \sqrt{Q/V}$ is the impact of selling the entire position, which can be large if the initial position is too big and/or the liquidity is too small. 
The leverage $\lambda(q)$ after $q$ shares have been sold is
\be
\lambda(q) = \frac{(Q-q)p(q)}{(Q-q)p(q)-L+C(q)},
\ee
where $p(q)$ is the price after selling $q$ shares.  Letting $x=q/Q$ be the fraction of assets that have been sold and $\lambda_0$ be the initial leverage before selling begins, this can be rewritten in the form
\be
\lambda(x) = \lambda_0  \left( \frac{(1-x)(1- {\cal I} \sqrt{x})}{1 - \lambda_0 {\cal I} \sqrt{x} \left(1 - x/3 \right)} \right).
\ee
It is then easy to show that:
\begin{itemize}
\item For small $x$, $\lambda(x) \approx \lambda_0 \left(1 + (\lambda_0 -1) {\cal I} \sqrt{x} \right)$, which is {\it larger} than $\lambda_0$ for $\lambda_0 > 1$, that is, whenever any leverage is used.  This means, rather paradoxically, that {\it when selling a leveraged position, the expected leverage under mark-to-market accounting always initially increases}.  

\item If $\lambda_0 {\cal I} < 3/2$ the leverage $\lambda(x)$ eventually reaches a maximum and decreases back to one for $x=1$. The crossover point $x^{*}$ where the leverage drops below its starting value can be computed by solving Eq.~(8) for $x$ with $\lambda(q) = \lambda_0$, which gives  
\be
x^* = \sqrt{\frac{1 - \sqrt{1 - \frac43 (\lambda_0-1)(3 - \lambda_0) {\cal I}^2}}{(2 - \lambda_0/3){\cal I}}}.
\ee
It is easy to show that $x^* < 1$ whenever $\lambda_0 {\cal I} < 3/2$.

\item {\it If $\lambda_0 {\cal I} > 3/2$ the leverage $\lambda(x)$ diverges during liquidation}.  The leverage diverges when the value of the position is equal to the liability, i.e. $qp(q) = L$.  This occurs when the denominator of Eq.~(9) becomes zero, which yields a cubic equation for the critical value $x = x_c$.  
If $x_c < 1$ then the asset manager goes bankrupt before being able to take his position to zero.  
\end{itemize}
Three representative deleveraging trajectories $\lambda(x)$ are illustrated in Fig.~1, together with the trajectory obtained in absence of market impact.  We assume a fixed starting mark-to-market leverage $\lambda_0 = 9$ and show three cases corresponding to different values of the overall market impact parameter $\mathcal{I}$.  For the two cases where the leverage is subcritical, i.e. with $\lambda_0 {\cal I} < 3/2$, the manager unwinds the position without bankruptcy.  However, due to the rise in leverage during the course of liquidation, he may get in serious trouble with his prime broker along the way.  For example, in the case where ${\cal I} = 0.15$ at its peak $\lambda(x)$ is more than twice its starting value.  

The case where the leverage is allowed to become supercritical is a disaster.  If $\lambda_0 {\cal I} > 3/2$, which for $\lambda_0 = 9$ implies ${\cal I} >  0.16$, the manager is trapped, and the likely outcome of attempting to deleverage is bankruptcy.  (By {\it bankruptcy} we mean that the position ends up being worth less than the money borrowed to finance it, so that the manager ends up owing a debt for that position. )

\begin{center}
\begin{figure}[h]
\begin{center}
\includegraphics[width=10cm]{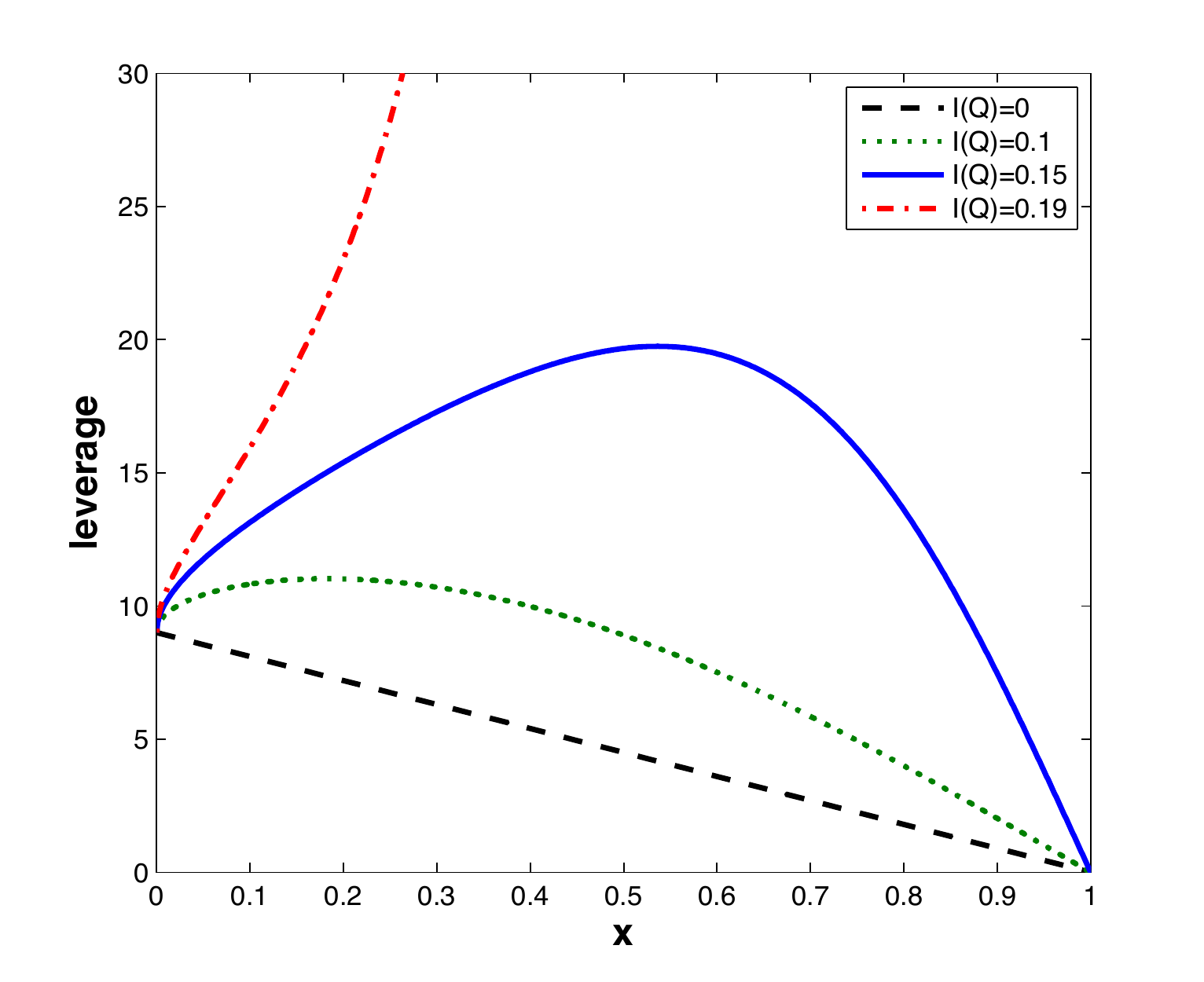}
\caption{\footnotesize{Possible deleveraging trajectories, showing the leverage $\lambda(x)$ based on mark-to-market accounting as a function of the fraction $x$ of the position that has been liquidated.  We hold the initial leverage $\lambda_0 = 9$ constant and show four trajectories for different values of the market impact parameter ${\cal I} = I(Q) = Y \sigma \sqrt{Q/V}$, i.e. ${\cal I} = 0$ (black dashed line, corresponding to the no-impact case) ${\cal I} = $ 0.1 (green dotted line),  0.15 (blue solid line), and 0.19 (red dotted-dashed line). If the market impact is too high the leverage diverges before the position can be liquidated, implying that the position is bankrupt. }} 
\label{fig1}
\end{center}
\end{figure}
\end{center}

\subsection{Leverage under impact-adjusted prices}

We now show how risk management is improved by impact-adjusted accounting.  This is done by simply using the average impact-adjusted valuation price $\tilde{p}$ in the formula for leverage, i.e.
\be
\tilde{\lambda}(q) = \frac{q\tilde{p}(q)}{q\tilde{p}(q) - L+C(q)}.
\ee
Here $0 \le q \le Q$ is the number of shares held at any given time along the way to entering position $Q$.  We define the impact adjusted price for position $q$ as the liquidation price if buying were to stop and the current position $q$ were to be sold.  Accordingly, when exiting a position we adopt the convention that the impact-adjusted price is based on complete liquidation of all $Q$ shares, i.e. we do not allow for the possibility of pausing along the way\footnote{
The square root law for market impact is inherently related to the market's memory \cite{Farmer11,Toth11b}.  Once liquidation begins the market has a memory -- the response of prices to each successive sale is smaller and smaller as $Q - q$ gets bigger and bigger.   An alternative definition of the impact-adjusted price while the position is being sold might be to assume a pause sufficiently long to break this memory, followed by subsequent liquidation of the remaining position $q$.  We have not adopted this because it requires an understanding of how impact decays in time, which we do not have, and in any case we do not believe this is necessary.}.

\begin{center}
\begin{figure}[h]
\begin{center}
\includegraphics[width=7cm]{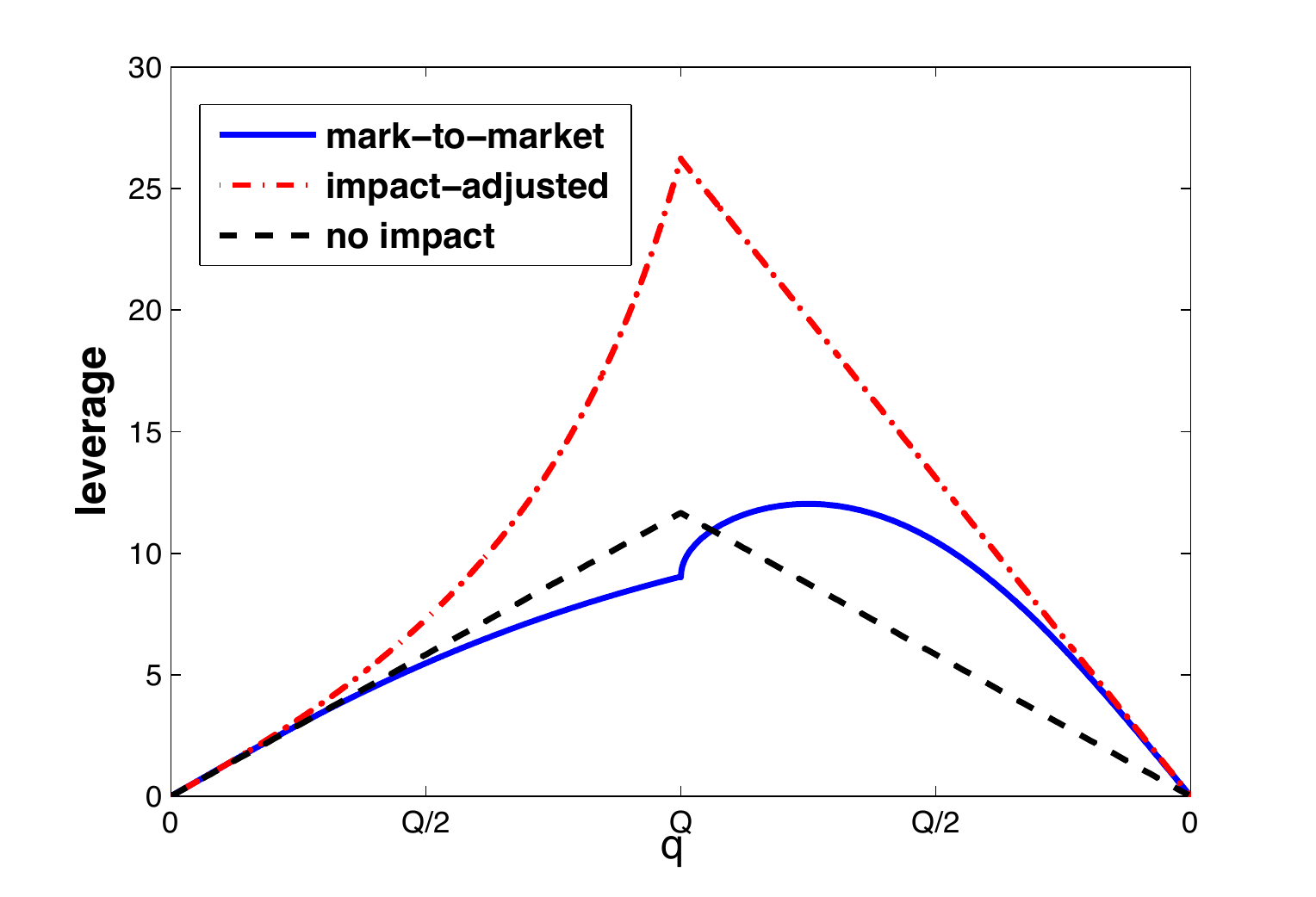}
\includegraphics[width=7cm]{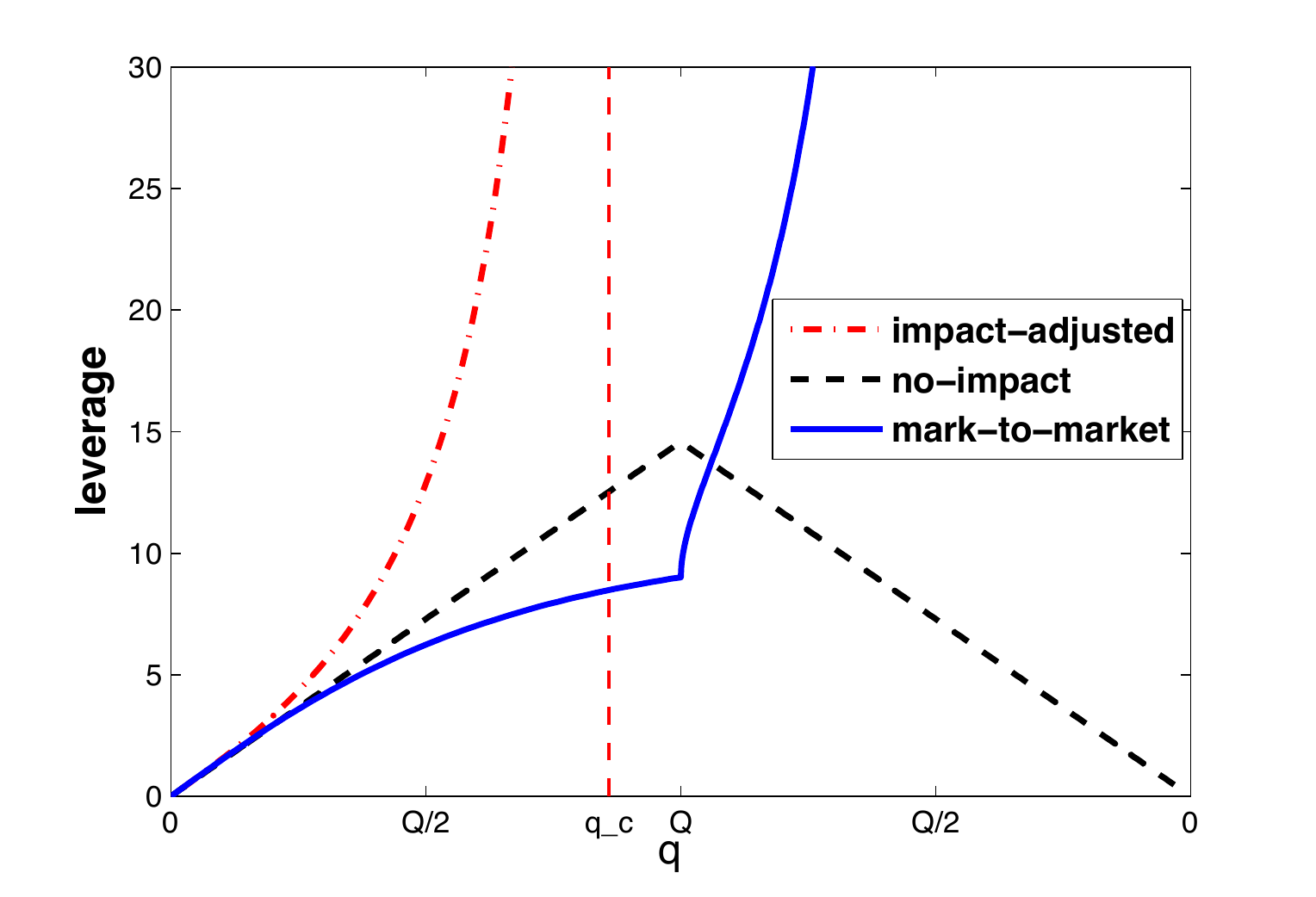}
\caption{\footnotesize{Leverage as a function of position size for first entering and then exiting a position. The position $q(t)$ varies from $0$ to $Q$ in the left half of each panel and from $Q$ to $0$ in the right half of each panel. Labels on the $x$ axes denote the number of shares held by the asset manager at the corresponding time. Three different measures are used for leverage.  The dashed black line shows what the leverage would be if there were no impact and the price didn't change; the solid blue line shows the leverage including impact under mark-to-market accounting, and the dotted-dashed red line shows the leverage using impact-adjusted valuation.  The left panel is a case in which $Q$ is small enough that the leverage never becomes critical; the right panel is a case where the leverage becomes super-critical.  In this case the impact-adjusted leverage diverges as the position is entered, warning the manager of the impending disaster. The dashed red vertical line shows the critical position $q_c$. }
}
\label{fig2}
\end{center}
\end{figure}
\end{center}

In figure \ref{fig2} we show how the leverage behaves when a manager first steadily assumes a position $0 \le q(t) \le Q$ and then steadily liquidates it.  We compare three different notions of leverage:  
\begin{itemize}
\item {\it No impact leverage} is represented by the dashed black line.  This is the leverage that would exist if the price remained constant (on average).  It rises and falls linearly\footnote{
The reason for linearity is that when the price is constant the denominator in Eq.~(\ref{before}) remains constant.  This is because changes in cash cancel changes in asset value.}
proportional to the position $q(t)$.
\item {\it Mark-to-market leverage} is represented by the solid blue line.  While the position is building it rises more slowly than linearly. This is because as the position is building impact causes the price to increase, lowering leverage and partially offsetting the increasing position size.  This is dangerous because it artificially overestimates profits and therefore depresses leverage.  When the position is exited, in contrast, the expected leverage initially shoots up.  In the subcritical case it eventually returns to zero, but in the super-critical case it diverges, indicating (too late) that the position is bankrupt.
\item
{\it Impact-adjusted} leverage is represented by the dashed red line.  It is always greater than either of the other two measures of leverage.  It is particularly useful in the super-critical case -- its rapid increase is a clear warning that a problem is developing, in contrast to the mark-to-market leverage.  A sensible manager would thus easily avoid bankruptcy by buying less and avoiding the critical regime.
\end{itemize}

\subsection{Taking noisy impact into account}

So far we have focused our attention on the expected impact, which can either be viewed as the average impact or as a median trajectory.  In this section we show how impact-adjusted accounting can be used to compute the probability of adverse price movements.  This improves on standard measures that fail to take impact into account and may dramatically underestimate the probability of bankruptcy in situations where impact is large.

To illustrate this we estimate the probability of bankruptcy for positions of varying leverage.  We make the simple assumption that the noisy component of impact is independent of the order being executed, and diffuses according to the volatility as the square root of time.  Under the (admittedly crude) approximation that background price movements are normally distributed 
we model individual realizations of price trajectories as a discrete random walk with time varying drift.  For convenience we measure the time $t$ in days. The evolution of the price during execution is given by
\be \label{noisyImpactEQ}
p(t+1) = p(t) + I(Q - q(t) - \delta q) - I(Q - q(t)) + p_0 \sigma n(t),
\ee
where $p(t)$ is the price at time $t$, $q(t)$ is the size of the position at time $t$, and $n(t)$ is IID gaussian noise.  The term $I(Q - q(t) - \delta q) - I(Q - q(t))$ is the additional increment of impact between day $t$ and day $t+1$, 
with $\delta q$ is the volume traded in a given day. The total time $T$ needed to off-load the position is $T=Q/\delta q$.

With this choice for the stochastic process we ensure that in absence of noise the price  follows the deterministic trajectory predicted by the expected market impact, and that the price in absence of market impact undergoes an unbiased discrete random walk.  For better risk analysis it is of course possible to use more sophisticated models of the background noise, incorporating factors such as clustered volatility, jump diffusions, or heavy tails as desired.  The simple model above is sufficient to illustrate the basic idea of how such risk analysis can be done.  

\begin{center}
\begin{figure}[h]
\begin{center}
\includegraphics[width=7cm]{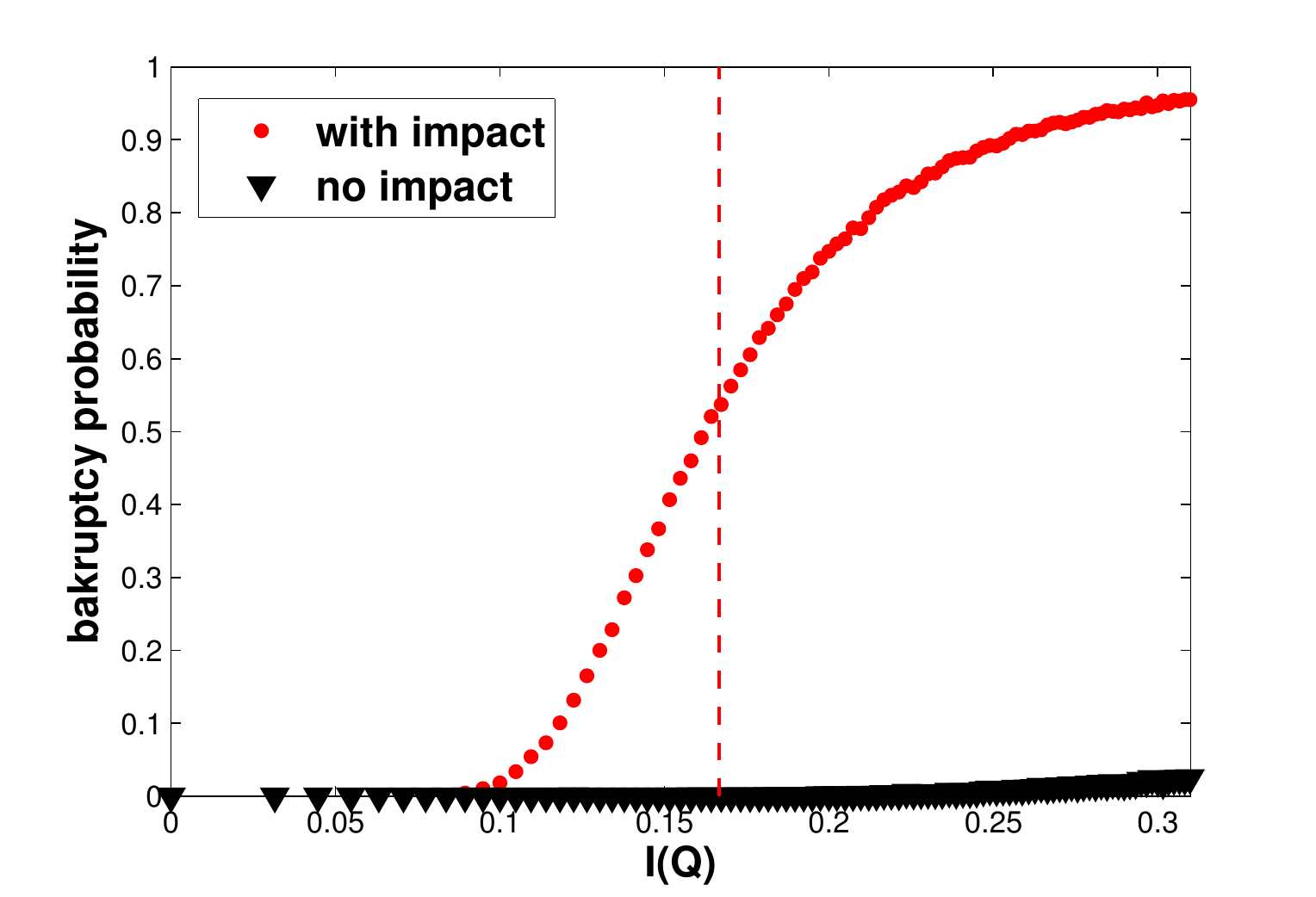}
\includegraphics[width=7cm]{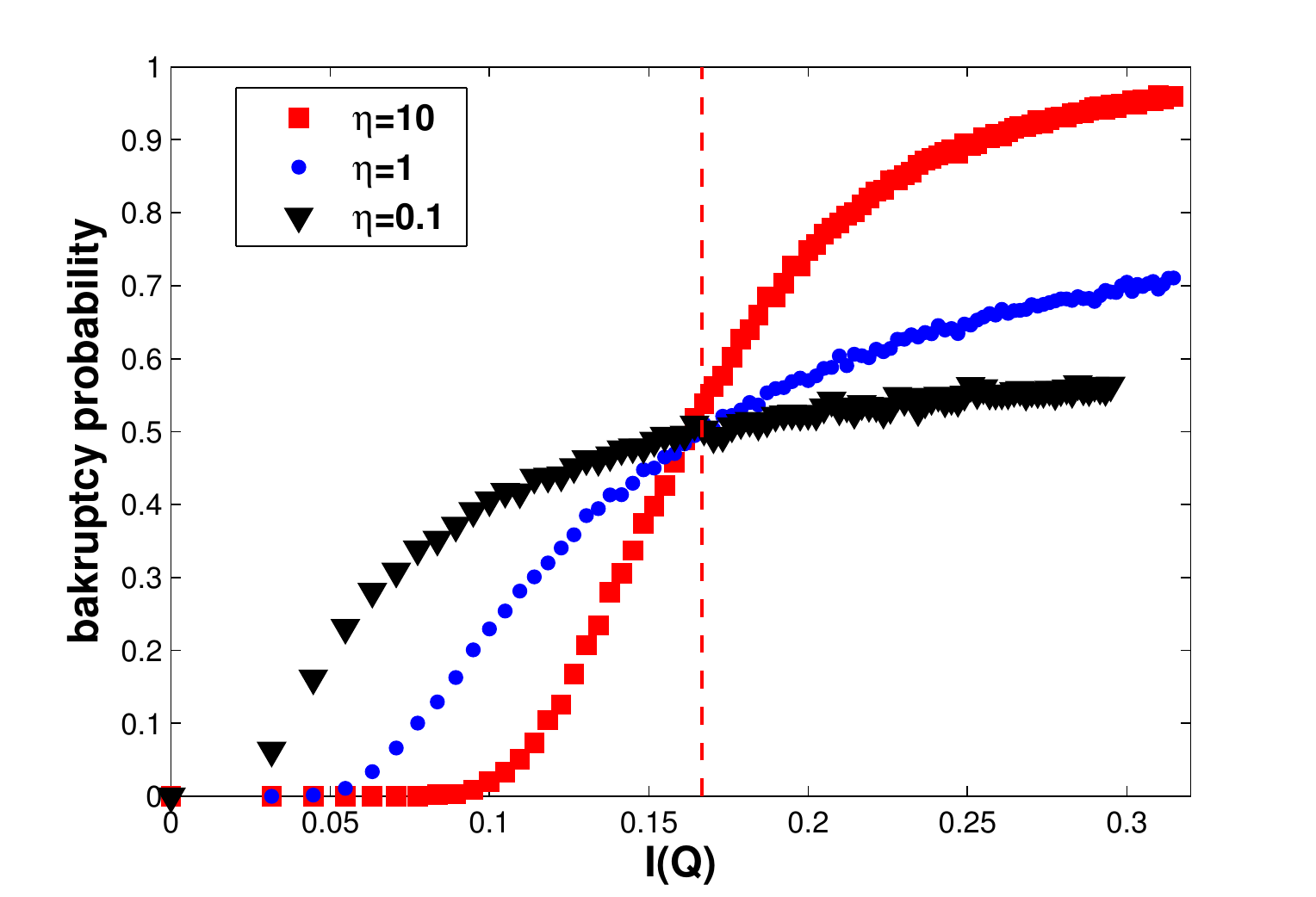}
\caption{\footnotesize{Bankruptcy probability as a function of the total impact $\mathcal{I}$. 
Probability of bankruptcy as a function of impact $I(Q)$.  In the left panel $\eta = Q/(VT) = 10$; $Q$ is varied from $0$ to $10^5$ while $T$ is varied to hold $\eta$ constant.  Red circles are bankruptcy probability with impact, and black triangles without impact.  The vertical dashed red line is the critical value $\mathcal{I}_c$.   The right panel is a similar plot for three different values of $\eta$.}}
\label{fig3}
\end{center}
\end{figure}
\end{center}

To assign probabilities for a given event, in this case bankruptcy, we simulate realizations of the noise process using Eq.~(\ref{noisyImpactEQ}), keeping $\sigma \sqrt{T}$ small enough that the probability that the price becomes negative can be neglected.  A typical result is shown in Fig.~\ref{fig3}.  Since the volatility of the final price scales as $\sigma \sqrt{T}$, whereas the average impact scales as $\sigma \sqrt{Q/V}$, the sharpness of the transition is determined by their ratio  $\eta \equiv Q/VT$.  Using $T = Q/\delta q$ we can write $\eta = \delta q/V$, making it clear that $\eta$ is an aggressivity parameter, often called the {\it participation rate}, measuring the fraction of daily volume used for trading.   In Fig.~\ref{fig3} we vary $Q$ and $T$ while keeping $\eta$ constant. 

The probabilities of bankruptcy are dramatically higher than they are without impact, and as expected, the transition is centered at the critical point $ \mathcal{I}_c = 3/(2 \lambda_0)$, which is independent of the volatility. The transition is sharp for aggressive trading schedules ($\eta > 1$) and is blurred as $\eta \to 0$.

During liquidation it is possible for the position to temporarily become bankrupt and then recover.  Whether or not a manager would be forced to default in such a situation will depend on her relationship with her creditors.  Forcing bankruptcy if it occurs anywhere along the liquidation path slightly raises the probability of bankruptcy, depending on the time for execution. 

\section{Does leverage diverge in realistic situations?}

We have shown the dangers of mark-to-market accounting for understanding leverage, but the skeptical reader may wonder whether such extreme situations actually occur in practice.  In this section we plug in some typical numbers and show that for large positions in illiquid stocks such problems are not uncommon.

\subsection{Impact and bid-ask spread}

We have so far computed impact using the estimated volatility and volume.  We now review results that connect these to the spread, which provide an alternative way 
to estimate the magnitude of liquidation effects, which might be more convenient in some circumstances.  

It is now well established empirically that the volatility is made up of two different effects: the size of the bid-ask spread $S$ on the one hand, and the number of transactions $\phi$ per unit time on the other. For liquid markets the volatility $\sigma_T$ on timescale $T$ can be written 
\be
\sigma_T = b S \sqrt{\phi T},
\ee
where $b \approx 0.6-0.9$ is a constant of order unity, that weakly depends on the market \cite{Wyart08}. Suppose that the typical volume at the best prices is $v$.  If one assumes as before that the order is executed in increments of $v$ shares, the total number of transactions needed to liquidate $Q$ shares is $N = Q/v$.   Similarly the total volume $V_T$ in a time $T$ is $V_T= v \phi T$. The above impact formula can therefore be rewritten as:\footnote{Note that the liquidation time $T$ drops out of the formula, which is one of the remarkable properties of the square-root impact law.}
\be\label{impactS}
I = Y \sigma_T \sqrt{\frac{Q}{V_T}} = Yb \, S  \sqrt{N}, \qquad N=\frac{Q}{v}.
\ee
This expression highlights the micro-structure origin of liquidity. As is intuitively clear, it is the spread $S$ and the available volume $v$ that determine the impact cost of a trade. 
The quantities $S$ and $v$ should again be estimated using moving averages using market data or broker quotes for OTC/illiquid markets. 

\subsection{Some examples}

Let us first give some orders of magnitude for stock markets. The daily volume of a typical stock is roughly $5 \times 10^{-3}$ of its market cap (see e.g. \cite{Lo00,Eisler06}), while its volatility is of the order $2 \%$ per day.  Suppose the portfolio to be liquidated owns $Q=5 \%$ of the market cap of a given stock. Taking $Y = 0.5$, the impact discount is  
\be
{I}(Q) \approx  2 \% \times \sqrt{\frac{0.05}{0.005}} \approx 6 \%.
\ee
A $6 \%$ hair-cut on the value of a portfolio of very liquid stocks is already quite large, and it is obviously much larger for less liquid/more volatile markets.  

Let us now turn to the question of the critical leverage $\lambda_c$ under mark-to-market accounting. From Section 3, the condition reads:
\be
\lambda_c \mathcal{I} = \frac{3}{2}.
\ee

Substituting the two expressions for $\mathcal{I} = I(Q)$ and rearranging gives
\begin{eqnarray}
\lambda_c & = & \frac{3}{2 Y \sigma} \sqrt{V/Q}\\& = & \frac{3}{2 Y b S \sqrt{N}}
\end{eqnarray}

To give a feeling for whether or not these conditions can be met, we present representative values for several different assets.  
For futures we assume $Q = V$, implying that it would take five days to trade out of the position with $\eta = 0.2$.  For stocks we assume $Q = 10V$, which assuming the same participation rate implies a position that would take 50 trading days to unwind.  Such positions might seem large, but they do occur for large funds; for instance, Warren Buffet was recently reported to have taken more than eight months to buy a $5.5\%$ share of IBM.   The results are given in Table~\ref{XXX}.   
 
\begin{table}
\begin{tabular}{| l c c c c c c c |} 
\hline 
Asset & $\sigma (daily)$ & $V (B \$)$ & $S (bp)$ & $v (M\$)$ & ${\cal I}_1^\star$ & ${\cal I}_2^\sharp$ & $\lambda_c$ \\ 
\hline 
BUND$^\dagger$& 0.4 \%  & 140 & 1.5 & 40 & 0.4\% & 0.7\% & $\sim$ 300 \\ 
SP500$^\dagger$ & 1.6\% & 150 & 2 & 10 & 1.6 \% & 2.1 \% & $\sim$ 100   \\ 
MSFT$^\diamondsuit$ & 2 \% & 1.25 & 3.7 & 1  & 6.3 \%  &  3.2 \% & $\sim$ 25   \\ 
AAPL$^\diamondsuit$ & 2.8 \% & 0.5 & 1.7 & 0.1 & 8.9\%  & 2.9 \% & $\sim$ 17  \\ 
KKR$^\heartsuit$ & 2.5 \% & 2$^\heartsuit$ & 14& 2.5$^\heartsuit$  & 7.9\%  & 9.4 \% & $\sim$ 16  \\ 
ClubMed$^\clubsuit$ & 4.3 \% & 1$^\clubsuit$ & 45 & 11$^\clubsuit$  & 13.5 \%  &  8.2 \% & $\sim$ 11   \\ 
CDS$^\flat$ & -- & -- & 10 \% & 10 & -- & 20\% & $\sim$ 7.5\\ 
\hline 
\end{tabular} 
\caption{\small{Numerical values of the different parameters entering the two alternative impact formulae given in Eqs. (\ref{impactQ}) and (\ref{impactS}) 
and the corresponding estimates of impact and critical leverage.  Except as otherwise noted, numbers are based on data for the first quarter of 2008. These are only rough orders of magnitude, intended for a qualitative discussion. 
$\star$: Impact $\mathcal{I}_1 = I(Q)$ based on volatility and volume, computed with Eq. (\ref{impactQ}), with $Y=1$ and $Q=V$ for 
futures and $Q=10V$ for stocks. This corresponds to a position of roughly $5 \%$ of the market capitalisation on stocks, and to a position equal to 3 Kerviels 
on the BUND.  $\sharp$: Impact computed with Eq. (\ref{impactS}), with $Y=1$, $b=0.6-0.9$ (depending on the market \cite{Wyart08}) and the same values of $Q$. 
$\dagger$: For futures, we refer to the nearest maturity; the numbers for the 10YUSNOTE are very similar to those for the BUND. Note that for liquid futures, the critical leverage level is very high (as expected). Still, 
a $1.5 \%$ liquidity hair-cut on a position on the SP500 is by no means negligible.  $\diamondsuit$: Large cap US stocks: In this case, $Q=10V$. Note that the two impact estimates are substantially different, 
with ${\cal I}_1 > {\cal I}_2$. This maybe due to the fact that the volume at the best quote, $v$, is highly skewed, i.e. the typical available volume is much smaller than the average volume. 
Furthermore, trades are usually only a fraction of the available volume. Therefore one expects $N > Q/v$. We have kept the more reliable formula  Eq. (\ref{impactQ}) to compute $\lambda_c$. 
$\heartsuit$ This is Krispy Kreme Doughnuts, a small cap US stock. $Q=10 V$ where $V$ is now in M \$ and $v$ in thousand \$. The numbers correspond to March 2012. 
$\clubsuit$ Club Med is a small cap French stock. $Q=10 V$, with $V$ is in M Euros. and $v$ in thousand Euros. The numbers correspond to 2002. 
$\flat$: For CDS on single names, these are OTC markets for which we only have estimates. Daily transactions are very patchy and their number is typically in the range $1 - 20$. 
We have chosen a reasonable value $N =Q/v = 10$, corresponding to a position of 10 to 100 M\$.  As expected, liquidity discount and potential deleveraging problems are very substantial here. 
}} 
\label{XXX} 
\end{table}

We see that for liquid futures, such as the BUND or SP500, the critical leverage is large enough that the phenomenon we discuss here is unlikely to ever occur.  As soon as we enter the world of equities, however, the situation looks quite different, whereas for OTC market the effect is certainly very real.

\section{Conclusion}

The above discussion underscores the need to value positions based on liquidation prices rather than mark-to-market prices.  For small, unleveraged positions in liquid markets there is no problem, but as soon as any of these conditions are violated, the problem can become severe.  As we have shown, standard valuations, which do nothing to take impact into account, can be wildly over-optimistic.

The solution that we have proposed accomplishes this goal by estimating liquidation prices based on recent advances in understanding market impact.  The procedures that we suggest have the key virtue of being extremely easy to implement.  They are based on quantities such as volatility, trading volume, or the spread, that are easy to measure.  Risk estimates can be computed for the typical expected behavior or for the probability of a loss of a given magnitude -- anything that can be done with standard risk measures can be easily replicated to take impact into account, with little additional effort.

The worst negative side-effects of mark-to-market valuations occur when leverage is used.  As we have shown here, when liquidity is low leverage can become critical.  By this we mean that as a position is being entered there is a critical value of the leverage $\lambda_c$ above which it becomes very likely that liquidation will result in bankruptcy, i.e. liquidation value less than money owed to creditors.  This does not require bad luck or unusual price fluctuations -- it is a nearly mechanical consequence of using too much leverage.   

Standard mark-to-market accounting gives no warning of this problem, in fact quite the opposite:  Impact raises prices as a position is purchased, causing leverage to be underestimated.  However, as a position is unwound the situation is reversed.  The impact of unwinding causes leverage to rise, and if the initial leverage is critical, the leverage becomes infinite and the position is bankrupt.  Under mark-to-market accounting this comes as a complete surprise.  Under impact-adjusted accounting, in contrast, the warning is clear.  As the critical point is approached the impact-adjusted leverage diverges, telling any sensible portfolio manager that it is time to stop buying.

The method of valuation that we propose here could potentially be used both by individual risk managers as well as by regulators.  Had such procedures been in place in the past, we believe that many previous disasters could have been avoided.  As demonstrated in the previous section, the values where leverage becomes critical are not unreasonable compared to those used before, such as the leverages of 50 - 100 used by LTCM in 1998, or 30-40 used by Lehman Brothers and other investment banks in 2008. 

However, one should worry about other potentially destabilizing feedback loops that our impact-adjusted valuation could trigger. For example, in a crisis situation, spreads and volatilities increase while
the liquidity of the market decreases, leading to a stronger discount on the asset valuation. But as was the case during the 2008 crisis, the write-down of the value of some books lead to further fire-sales, fueling more panic. So it is important to estimate the parameters entering the impact formula (volatility, spread and available volumes) using a slow moving average to avoid any over-reaction to temporary liquidity draughts.  

A key point underlying our discussion here is that market impact occurs for both informed and uninformed trades.   Empirical studies make it clear that temporary market impact occurs even if trades are made for reasons, such as hedging or liquidity, that have nothing to do with underlying fundamentals.   This should not be surprising:  Typically the counterparty has no way of knowing whether the opposite side of the trade is ``informed" or ``uninformed"\footnote{
To be clear, there are two kinds of impact.  The first is due to correctly anticipating price movements.  On the timescales for completing a trade this is typically small.  The second is due to influencing prices.  Our point is that if my counterparty is anonymous I have no way of knowing how informed she is, and therefore must react in a generic manner.}

The failure of mark-to-market accounting can thus be viewed as a failure of the theory of efficient markets, or at the very least the need to take a liberal view of what it means. The fact that prices can change substantially due to random events that have nothing to do with fundamentals reflects a failure of prices to provide accurate valuations.  Alternatively, one can take a generous view of what the word ``accurate" means, as Fisher Black did when he famously said, ``{\it we might define an efficient market as one in which price is within a factor of 2 of value $\ldots$ By this definition I think almost all markets are efficient almost all of the time.  `Almost all' means at least $90\%$}" \cite{Black86}. 
 
The failure of marginal prices as a useful means of valuation is part of an emerging view of markets as dynamic, endogenously driven and self-referential \cite{Sornette05,JP_Risk}, as suggested long ago by Keynes \cite{Keynes} and more recently by Soros \cite{Soros08}. For example, recent studies suggest that exogenous news play a minor role in explaining major price jumps \cite{Joulin08}, while self-referential feedback effects are strong \cite{Sornette12}. Market prices 
are molded and shaped by trading, just as trading is molded and shaped by prices, with intricate and sometimes destabilising feedback. Because the 
liquidity of markets is so low, the impact of trades is essential to understand why prices move \cite{Bouchaud08b}. 

\section*{Acknowledgments}
This work was supported by the National Science Foundation under grant 0965673, by the European Union Seventh Framework Programme FP7/2007-2013 under grant agreement CRISIS-ICT-2011-288501, and by the Sloan Foundation. JPB acknowledges important discussions with X. Brokmann, J. Kockelkoren and B. Toth.
\bibliography{jdf}{}
\bibliographystyle{ieeetr}

\end{document}